\documentclass[apl,aip,reprint]{revtex4-1} 

\usepackage[dvips]{graphicx}
\usepackage{bm}
\usepackage{latexsym} 
\usepackage{amsmath}
\usepackage{amssymb}

\newcommand{\ui}{{\rm i}}

\newcommand{\bmq}{{\bm q}}

\newcommand{\bmk}{{\bm k}}

\newcommand{\bmK}{{\bm K}}

\newcommand{\kB}{k_{\rm B}}

\begin{document}

\title{
Gigantic enhancement of spin Seebeck effect by phonon drag 
}

\author{
Hiroto Adachi$^{1,2}$ 
}
\email[]{hiroto.adachi@gmail.com} 

\author{
Ken-ichi Uchida$^{3}$ 
}

\author{
Eiji Saitoh$^{3,1,2,4}$
}

\author{
Jun-ichiro Ohe$^{1,2}$
} 

\author{
Saburo Takahashi$^{3,2}$
}

\author{
Sadamichi Maekawa$^{1,2}$
}

\affiliation{ 
$^{1}$ 
Advanced Science Research Center, Japan Atomic Energy Agency, 
Tokai 319-1195, Japan\\
$^{2}$ 
CREST, Japan Science and Technology Agency, Sanbancho, Tokyo 102-0075, Japan\\
$^{3}$ 
Institute for Materials Research, Tohoku University, 
Sendai 980-8755, Japan\\
$^{4}$ 
PRESTO, Japan Science and Technology Agency, Sanbancho, 
Tokyo 102-0075, Japan  
}

\date{\today}

\begin{abstract} 

We investigate both theoretically and experimentally 
a gigantic enhancement of the spin Seebeck effect 
in a prototypical magnet LaY$_2$Fe$_5$O$_{12}$ at low temperatures. 
Our theoretical analysis sheds light on the important role of 
phonons; 
the spin Seebeck effect is enormously enhanced by nonequilibrium phonons 
that drag the low-lying spin excitations. 
We further argue that this scenario gives a clue to understand 
the observation of the spin Seebeck effect that is unaccompanied 
by a global spin current, and predict that the substrate 
condition affects the observed signal. 

\end{abstract}



\maketitle 

When a temperature gradient is applied to a ferromagnet, 
a force is induced acting on electrons' spin to drive spin currents. 
This phenomenon termed the spin Seebeck 
effect (SSE)~\cite{Uchida08,Jaworski10,Uchida10} has recently 
drawn tremendous attention as a new source of 
spin currents 
needed for future spin-based electronics.~\cite{Zutic04,Maekawa06} 
SSE is now established as an universal aspect of 
ferromagnetic materials as it has been observed in a variety of 
materials ranging from a metallic ferromagnet~\cite{Uchida08} 
Ni$_{81}$Fe$_{19}$ and a semiconducting ferromagnet~\cite{Jaworski10} GaMnAs 
to an insulating magnet~\cite{Uchida10} LaY$_2$Fe$_5$O$_{12}$. 
Besides its impact on the technological application, 
SSE offers a number of new topics on the interplay of heat and 
spin currents,~\cite{Slonczewski10,Slachter10} 
and it triggered the emergence of the new field named 
``spin caloritronics''~\cite{SpinCalo} in the rapidly-growing spintronics community.

A mystery concerning SSE was 
how conduction electrons can sustain the 
spin voltage over so long range of several millimeters 
in spite of the short conduction electrons' spin-flip 
diffusion length $\lambda_{\rm sf}$, which is typically 
of several tens nanometers. 
This problem has recently been resolved by a series of experiments 
on spin currents using magnetic insulators. 
A recent experiment on the electric signal transmission through 
a magnetic insulator~\cite{Kajiwara10} highlights the role of 
the low-lying magnetic excitation of localized spins, i.e., 
magnons, by demonstrating that 
magnons transmit the spin current over a long distance of several millimeters. 
A subsequent experiment on SSE for a magnetic insulator~\cite{Uchida10} 
LaY$_2$Fe$_5$O$_{12}$ confirmed that the 
magnon-based scenario can explain the SSE experiment at room temperature, 
since the length scale associated with magnons $\gg \lambda_{\rm sf}$. 
However, a new issue on SSE was brought by a very recent experiment on 
a ferromagnetic semiconductor~\cite{Jaworski10} GaMnAs, where it was demonstrated, 
by cutting the magnetic coupling in GaMnAs while keeping the thermal 
contact, that SSE can be observed even in the 
absence of global spin current flowing through GaMnAs. 
Obviously, the scenario of magnon-mediated SSE~\cite{Xiao10,Adachi10} fails to 
explain the experiment, showing 
that the full understanding of SSE has not yet been reached. 

\begin{figure}[b] 
  \begin{center}
    \scalebox{0.9}[0.9]{\includegraphics{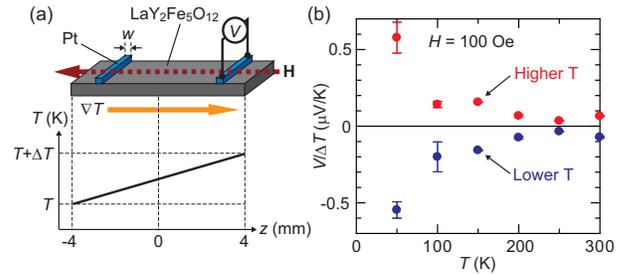}}
  \end{center}
\caption{ (Color online) Gigantic enhancement of SSE in LaY$_2$Fe$_5$O$_{12}$ 
at low temperatures. 
(a) Schematic illustration of the LaY$_2$Fe$_5$O$_{12}$/Pt sample 
and the temperature profile along the $z$ direction. 
Here ${\bm H}$ denotes an external magnetic field (with magnitude $H$). 
The sample comprises a LaY$_2$Fe$_5$O$_{12}$ film with 
$8 \times 4\,\textrm{mm}^2$ rectangular shape and two separated Pt wires 
with the width $w$ 
attached to the LaY$_2$Fe$_5$O$_{12}$ surface at the interval of 5.6 mm. 
(b) $T$ dependence of $V/\Delta T$ at $H=100\,\textrm{Oe}$. } 
\label{fig1_SSEphonon}
\end{figure}

For a deep understanding of the physics behind SSE, 
we here explore the low-temperature behavior of SSE in an insulating 
magnet LaY$_2$Fe$_5$O$_{12}$. 
Figure~\ref{fig1_SSEphonon} shows a schematic illustration of 
our device structure.~\cite{suppl01} 
An in-plane external magnetic field ${\bm H}$ and a uniform temperature gradient 
${\bm \nabla} T$ were applied along the z direction [see FIG.~\ref{fig1_SSEphonon}~(a)]. 
The ${\bm \nabla} T$ generates a spin voltage across 
the LaY$_2$Fe$_5$O$_{12}$/Pt interface, 
and injects (ejects) a spin current $I_s$ into (from) the Pt wire. 
In the Pt wire, a part of the injected/ejected $I_s$ is converted 
into a charge voltage through the so-called inverse spin-Hall 
effect (ISHE):~\cite{Saitoh06} 
\begin{eqnarray}
  V_{\rm ISHE} &=& \Theta_H (|e|I_s)(\rho/w), 
  \label{Eq:ISHE01}
\end{eqnarray}
where $|e|$, $\Theta_H$, $\rho$ and $w$ are the absolute value of 
electron charge, spin-Hall angle, resistivity and width of the 
Pt wire, respectively. 
Therefore, the $\nabla T$-driven spin injection, or SSE, is electrically detectable. 
In FIG.~\ref{fig1_SSEphonon}~(b), we show the temperature ($T$) 
dependence of $V_{\rm ISHE}/\Delta T$ at 
$H=100 \, \textrm{Oe}$, measured when the Pt wires are attached to the lower- 
and higher-temperature ends of the LaY$_2$Fe$_5$O$_{12}$ layer, respectively. 
The sign of $V/\Delta T$ is reversed between these Pt wires, 
a situation consistent with SSE-induced ISHE. 
Notable is that, at $T=50 \, \textrm{K}$, the magnitude of $V/\Delta T$ is 
dramatically enhanced.

\begin{figure}[t] 
  \begin{center} 
    \scalebox{0.6}[0.6]{\includegraphics{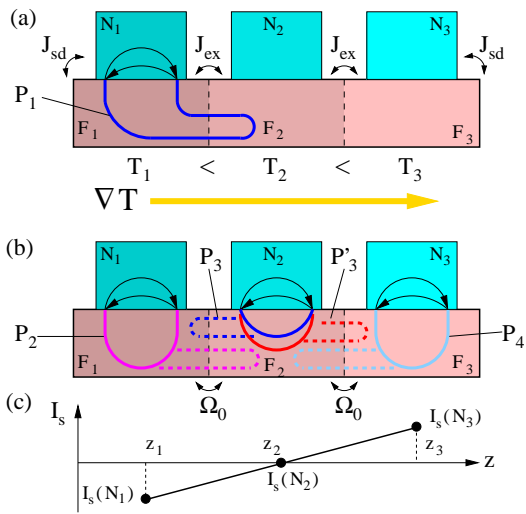}} 
  \end{center}
\caption{(Color online) 
Diagrammatic representation of the thermal spin injection process. 
(a) Magnon-mediated SSE. Here the system is composed of 
ferromagnet ($F$, in the experiment LaY$_2$Fe$_5$O$_{12}$) 
and nonmagnetic metals ($N$, in the experiment Pt), which are 
divided into three temperature domains of $F_1/N_1$, $F_2/N_2$, and 
$F_3/N_3$ with their local temperatures of $T_1$, $T_2$, and $T_3$. 
The thin solid lines with arrows 
(bold lines without arrows) are electron (magnon) propagators. 
Here, $J_{\rm sd}$ ($J_{\rm ex}$) is the strength of the $s$-$d$ coupling 
at the $F/N$ interface (the exchange coupling in 
$F$). 
(b) Phonon-dragged SSE where the dashed lines are phonon propagators. 
The process $P_4$ injects the spin current 
with the same magnitude as (but opposite sign to) the process $P_2$ 
due to the relation $T_1-T_2 = -(T_3-T_2)$, while no spin current 
is injected into $N_2$ because of the cancellation between 
the two relevant processes 
$P_3$ and $P'_3$. Here, $\Omega_0=\sqrt{K_{\rm ph}/M_{\rm ion}}$ 
with the ion mass $M_{\rm ion}$ and the elastic constant $K_{\rm ph}$ in $F$. 
(c) Calculated spatial dependence of the spin current injected into $N_i$ 
$(i=1,2,3)$. 
}
\label{fig2_SSEphonon} 
\end{figure}

A simple scenario of the magnon-mediated SSE~\cite{Xiao10,Adachi10} 
[FIG.~\ref{fig2_SSEphonon}~(a)] 
is unable to explain the observed low-$T$ enhancement. 
If such a scenario could explain the experiment, 
the low-$T$ enhancement of $V_{\rm ISHE}$ would come from 
either the enhancement of the spin-Hall angle $\Theta_H$ or 
that of the magnon lifetime; otherwise $I_s$ due to magnon-mediated SSE 
is a monotonic increasing function of $T$ 
as discussed below. 
From the $T$-dependence of the spin-Hall 
conductivity,~\cite{Vila07} we conclude that there is no enhancement 
of $\Theta_H$ at low $T$. 
While the possibility of the enhancement of 
magnon lifetime is not conclusively excluded, 
judging from the ferromagnetic resonance linewidth in 
Y$_3$Fe$_5$O$_{12}$~\cite{Vittoria85} as a measure of the inverse 
magnon lifetime, it does not seem to be the case. 
Therefore, we need a new mechanism to account for the 
observed low-$T$ enhancement of SSE.

Here, the gigantic enhancement of SSE for LaY$_2$Fe$_5$O$_{12}$ 
below room temperature is analyzed in the light of 
phonon-drag mechanism.~\cite{Gurevich46,Blatt76,Lifshitz} 
Back in 1946 in the context of thermoelectrics, 
Gurevich pointed out~\cite{Gurevich46} 
that the thermopower can be generated by a stream of 
phonons driven by the temperature gradient, 
which then drag electrons and cause their convection. 
This idea, known nowadays as phonon-drag mechanism, has been 
established~\cite{Blatt76,Lifshitz} as a principal mechanism causing 
the low-$T$ enhancement~\cite{phonon-drag-exp} 
of the thermopower. 
Because SSE is a spin counterpart of the Seebeck effect, 
it is natural to expect that a similar physics underlies SSE. 
It is this approach 
that we adopt in the present work.

\begin{figure}[t] 
  \begin{center}
    \scalebox{0.6}[0.6]{\includegraphics{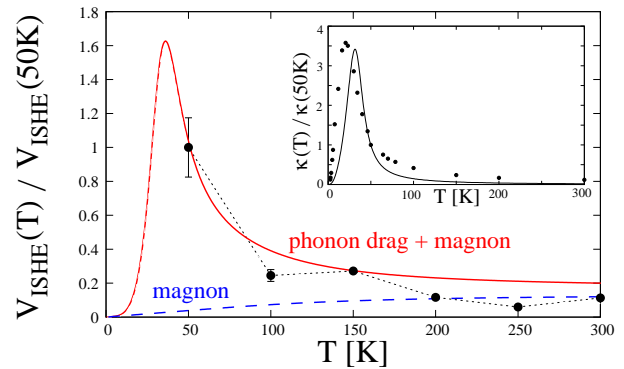}} 
  \end{center}
  \caption{ (Color online) 
    Comparison of experimental and theoretical SSE signal. 
    Solid circles: experimental spin Seebeck data for LaY$_2$Fe$_5$O$_{12}$ 
    (Dotted line is a guide to the eye). 
    The solid curve: calculated $T$-dependence of $V_{\rm ISHE}$ 
    due to the sum of the phonon-dragged SSE and the magnon-mediated SSE. 
    The dashed curve: calculated $T$-dependence of $V_{\rm ISHE}$ 
    due only to the magnon-mediated SSE. 
    We have assumed $T$-independent $\Theta_H$ and $\alpha$, 
    and used $T_D=565 \, {\rm K}$~\cite{Slack71} and $T_M=560 \, {\rm K}$. 
    The data are normalized by its value at $50$ K except that 
    the result for magnon-mediated SSE is plotted to reproduce 
    the room-temperature signal. 
    Inset: experimental thermal conductivity $\kappa$ for Y$_3$Fe$_5$O$_{12}$ 
    taken from Ref.~\onlinecite{Slack71} (solid circles) and 
    the result of the fit (solid curve) using Eq.~(\ref{Eq:kappa}). 
  }
\label{fig3_SSEphonon}
\end{figure}

Our theoretical analysis starts from considering the model 
shown in FIG.~\ref{fig2_SSEphonon}~(a). 
The key point in our model is that the temperature gradient $\nabla T$ 
is applied over the ferromagnet, but there is locally no temperature 
difference between the ferromagnet ($F$) and the attached 
nonmagnetic metals ($N$). 
We assume that each temperature domain is initially in local thermal equilibrium, 
then we switch on the interactions among the domains and calculate the 
nonequilibrium dynamics of spin density in $N$.

The central quantity that characterizes SSE is the spin current 
$I_s$ injected into $N$ (in experiment Pt), since it is proportional to 
the experimentally-detectable electric voltage via ISHE [Eq.~(\ref{Eq:ISHE01})]. 
Following Ref.~\onlinecite{Adachi10}, 
the spin current $I_s^{\rm mag}(N_1)$ injected into $N_1$ 
due to the magnon-mediated SSE~\cite{Xiao10,Adachi10} 
is calculated  as 
\begin{eqnarray}
  I_s^{\rm mag}(N_1)/\Delta T &=& 
  \left( \frac{P}{\alpha} \right) 
  \int_{0}^{T_M/T} 
  ds \frac{  ({T}/{T_M}) \, s^2} 
  {4 \sinh^2(\tfrac{s}{2})}, 
  \label{Eq:Is_nonloc01} 
\end{eqnarray}
where $\alpha$ is the Gilbert damping constant, 
$T_M$ is the characteristic temperature corresponding to 
the magnon high-energy cutoff, 
and $P$ is a nearly-temperature-independent coefficient.~\cite{com01} 
Equation~(\ref{Eq:Is_nonloc01}) means that the magnon-mediated SSE cannot 
explain the low-$T$ enhancement of the signal 
(the dashed curve in FIG.~\ref{fig3_SSEphonon}).

\begin{figure}[t] 
  \begin{center}
    \scalebox{0.6}[0.6]{\includegraphics{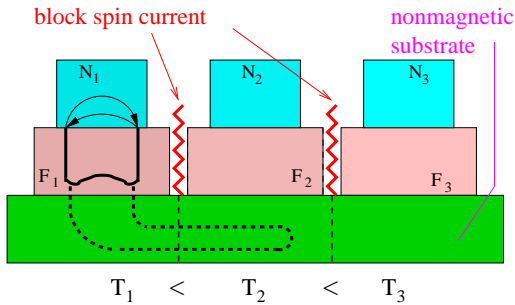}} 
  \end{center}
  \caption{ (Color online) 
    Schematic illustration of SSE unaccompanied by a 
    global spin current. The phonon-drag process which explains the 
    experiment~\cite{Jaworski10} is shown. The meaning of each 
    line (propagator) is the same 
    as in FIG.~\ref{fig2_SSEphonon}. 
  }
\label{fig4_SSEphonon}
\end{figure}

Now we proceed to a detailed analysis of the $T$-dependence of SSE 
in terms of the phonon-drag mechanism. 
The Feynman diagram for the phonon-drag process in the present situation 
is shown in FIG.~\ref{fig2_SSEphonon}~(b), 
where the phonons feel the temperature difference between $F_1$ and $F_2$, 
and drag magnons through the magnon-phonon interaction. 
Since the nonequilibrium phonons affect the magnon dynamics, 
this process injects spin current into $N_1$. 
The important point is that the spin current $I_s^{\rm drag}$ 
injected in this process becomes proportional to the phonon lifetime 
$\tau_{\rm ph}$ as~\cite{suppl01} 
\begin{eqnarray}
  I_s^{\rm drag}(N_1)/ \Delta T 
  &=&
  P' \tau_{\rm ph} {\cal B}_1 {\cal B}_2, 
  \label{Eq:Is_phonon-drag01}
\end{eqnarray}
where 
$ {\cal B}_1 = 
\frac{(T/T_D)^5}{4\pi^3}  \int_0^{T_D/T} 
d u \frac{u^6}{\sinh^2 (u/2)}$ 
and 
$ {\cal B}_2  = 
\frac{(T/T_M)^{9/2} } {4 \pi^2} 
(\frac{\kB T_M \tau_{\rm sf}}{\hbar})^3 
\int_0^{T_M/T} d v \frac{v^{7/2}}{\tanh(v/2)} $ 
with the Debye temperature $T_D$, 
and 
$P'$ is a nearly-temperature-independent coefficient.~\cite{com02} 
Since $\tau_{\rm ph}$ in a high-purity specimen is known to 
increase steeply at low $T$ because of the rapid suppression of 
umklapp scattering,~\cite{Ashcroft76} 
it leads to the drastic enhancement of the phonon-dragged SSE. 
In our analysis, the $T$-dependence of $\tau_{\rm ph}$ is extracted 
from the thermal conductivity data for Y$_3$Fe$_5$O$_{12}$~\cite{Slack71} 
(see the inset of FIG.~\ref{fig3_SSEphonon}) using~\cite{Ashcroft76,com03} 
\begin{eqnarray}
\kappa (T) &=&
(1/3) v_{\rm ph}^2 C_{\rm ph} (T) \tau_{\rm ph} (T), 
\label{Eq:kappa}
\end{eqnarray} 
where 
$v_{\rm ph}$ is the phonon velocity, and 
$C_{\rm ph}(T) = 9N_D \kB ( {T}/{T_D} )^3 
\int_0^{T_D/T} dw \frac{w^3}{4 \sinh^2(w/2)} $ 
is the phonon specific heat with the number of phonon modes $N_D$. 
After getting the information on $\tau_{\rm ph}(T)$, we calculate 
the $T$-dependence of $V_{\rm ISHE}$ resulting from the phonon-dragged SSE. 
The result, plotted in FIG.~\ref{fig3_SSEphonon} (the solid curve), 
shows an excellent description of the low-$T$ enhancement of SSE.~\cite{com04} 
Our analysis demonstrates that the phonon-drag mechanism is of 
crucial importance to understand SSE below the room temperature.

Finally, we show in FIG.~\ref{fig4_SSEphonon} our interpretation 
on the observation of SSE that is unaccompanied by a global 
spin current,~\cite{Jaworski10} 
where the heat is carried by phonons through the nonmagnetic 
substrate while the spin is injected locally at the $F/N$ interface. 
This interpretation is reinforced when we recall that the magnitude 
of the spin Seebeck signal is enhanced with decreasing $T$ 
even well below the Curie temperature, 
whose tendency is consistent with 
the phonon-drag mechanism as is seen in FIG.~\ref{fig3_SSEphonon}. 
Furthermore, the fact that the experiment was done 
below the room temperature supports the the phonon-drag-based 
scenario, since the phonon-drag process becomes more effective at low $T$ 
as emphasized in the previous paragraph. 
All these considerations strongly support 
that the SSE experiment for GaMnAs can be 
interpreted in terms of the phonon-drag mechanism, and results in 
a prediction that the substrate condition affects the observed signal. 
Our demonstration opens a new route to control spin currents 
by means of phonons and stimulates further progresses in 
spin caloritronics. 

This work was supported by a Grant-in-Aid for Scientific Research 
in Priority Area 
'Creation and control of spin current' (19048009, 19048028), 
a Grant-in-Aid for Scientific Research A (21244058), 
the Global COE for the 'Materials Integration International Centre 
of Education and Research', 
a Grant-in-Aid for Young Scientists (No. 22740210) 
all from MEXT, Japan, 
a Grant for Industrial Technology Research from NEDO, Japan, 
Fundamental Research Grants from CREST-JST, PRESTO-JST, TRF, and TUIAREO, Japan. 



\begin{thebibliography}{99}
\bibitem{com01} 
  $P=(\kB \tau_{\rm sf}/8 \pi^5 \hbar^3)\times L$ 
  with $L= 0.1 \times 
  N_{\rm int} (J_{\rm sd}^2 S_0 ) \chi_N (a/\lambda_{\rm sf})^3 (a_S/\Lambda)$, 
  where $N_{\rm int}$ and $J_{\rm sd}$ are the number of localized spins 
  and the strength of the $s$-$d$ exchange coupling at the $N_1$-$F_1$ interface; 
  $\chi_N$, $\tau_{\rm sf}$, and $\lambda_{\rm sf}$ are the paramagnetic susceptibility, 
  the spin relaxation time, and the spin diffusion length in $N_1$; 
  $\Lambda$ and $a_S$ are the dimension and 
  lattice constant of the effective block spin of $F_1$.
\bibitem{com03} 
  For the $T$-dependence of $\tau_{\rm ph}$, we introduce a model 
  $\tau_{\rm ph}(T)/\tau_{\rm ph}(T=0) = [ 1+ \frac{1}{a_0} e^{-b_0(T_D/T)} ]^{-1} 
  [1+ \frac{1}{c_0}(T/T_D) ]^{-1}$ 
  with fit parameters $(a_0,b_0,c_0) =  (2 \times 10^{-4},0.55,0.01)$, 
  where the first term represents the low-$T$ enhancement (ref.~\onlinecite{Ashcroft76}) 
  $\tau_{\rm ph} \sim e^{b_0T_D/T}$ originating from 
  the steep suppression of the umklapp scattering 
  at $T \sim b_0 T_D$, 
  while the second term captures the high-$T$ 
  behavior (ref.~\onlinecite{Ashcroft76}) $\tau_{\rm ph} \sim 1/T$.
\bibitem{com04}
  Note that theoretically the signal due to the phonon-dragged 
  SSE has a close-to-linear spatial dependence [see FIG.~\ref{fig2_SSEphonon} (c)] 
  as is observed in experiments, 
  the reason of which is obvious from FIG.~\ref{fig2_SSEphonon} (b). 
  In addition, the length scale associated with the phonon-dragged SSE 
  is set by that related to the energy conservation, such that it can 
  be as long as several millimeters.
\end{thebibliography}

\appendix
\section*{Supplemental Material} 

\subsection{Experimental details} 
The single-crystal LaY$_2$Fe$_5$O$_{12}$ (111) film with the thickness of 
$3.9\,\mu \textrm{m}$ was grown on a Gd$_3$Ga$_5$O$_{12}$ (111) substrate by 
liquid phase epitaxy. The 15-nm-thick Pt wires were then sputtered in an Ar 
atmosphere. The length and width of each Pt wire are 4 mm and 0.1 mm, respectively. 
The temperatures of the lower- and 
higher-temperature ends of the sample were respectively stabilized to $T$ and 
$T+\Delta T$, where $T$ was controlled in the range of 300-50 K by means of 
a closed-cycle helium refrigerator.

\subsection{Derivation of Eq.~(3)} 
Following Ref.~[S1], the spin current $I_s(N_i,t)$ injected into 
the nonmagnetic metal $N_i$ ($i=1,2,3$) 
is calculated as 
\begin{equation}
  I_s(N_i,t) 
  = -\sum_{\bmq,\bmk} 
  \frac{4{\cal J}^{\bmk- \bmq}_{\rm sd} \sqrt{S_0} } 
       {\sqrt{2 N_F N_N} \hbar} 
       {\rm Re} C^{<}_{\bmk,\bmq}(t,t), 
       \label{Eq:I_s00} 
       \tag{S1} 
\end{equation}
where $N_F$ ($N_N$) is the number of lattice sites 
in $F$ ($N$), $S_0$ is the size of the localized spins in $F$, 
and ${\cal J}_{\rm sd}^{\bmk+\bmq}$ is the Fourier transform of the 
$s$-$d$ interaction at the $F/N$ interface. 
Here, 
$C^{<}_{\bmk,\bmq}(t,t') = - \ui \langle  a^+_\bmq(t') s^-_\bmk(t) \rangle $ 
measures the correlation between 
the magnon operator $a_\bmq^+$ and the spin-density operator 
$s^-_\bmk= (s^x_\bmk- \ui s^y_\bmk)/2$. 
Note that the time dependence of $I_s(N_i,t)$ vanishes in the steady state 
and it is hereafter discarded. 
Introducing the frequency representation 
$C^{<}_{\bmk,\bmq}(t-t') = 
\int_{-\infty}^\infty \frac{d \omega}{2 \pi} 
{C}^{<}_{\bmq,\bmk}(\omega) e^{- \ui \omega (t-t')}$ 
and adopting the representation~[S2] 
$\check{C} 
= \left({{C^{R}, C^{K}} \atop {0 \;\;\;  ,C^{A}}} \right)$
as well as using the relation $C^{<}= \frac{1}{2} [C^{K}- C^{R} + C^{A}]$, 
we obtain 
\begin{equation}
  I_s(N_1) =   \sum_{\bmq,\bmk} 
  \frac{-2{\cal J}^{\bmk-\bmq}_{\rm sd} \sqrt{S_0} }
  {\sqrt{2 N_F N_N} \hbar} 
  \int_{-\infty}^\infty \frac{d \omega}{2 \pi} 
  {\rm Re} C^{K}_{\bmk,\bmq}(\omega) 
  \label{Eq:I_s01}
  \tag{S2} 
\end{equation}
for the spin current $I_s(N_1)$ in the steady state. 

When we introduce a renormalized magnon propagator $\delta \check{X}_\bmq(\omega)$, 
the interface correlation $\check{C}$ appearing in Eq.~(\ref{Eq:I_s01}) 
is generally expressed as 
\begin{equation}
  \check{C}_{\bmk,\bmq} (\omega) = 
  \frac{{\cal J}^{\bmk-\bmq}_{\rm sd} \sqrt{S_0} }{\sqrt{N_N N_F} \hbar}
  \check{\chi}_\bmk(\omega) \delta \check{X}_\bmq(\omega), 
  \label{Eq:C-func01}
  \tag{S3}
\end{equation} 
where 
$\check{\chi}_\bmk(\omega) 
=
\left( { \chi_\bmk^R (\omega), \atop 0,} 
     { \chi_\bmk^K (\omega) 
       \atop \chi_\bmk^A (\omega)} 
     \right) $ 
is the spin-density propagator 
satisfying the local equilibrium condition: 
\begin{equation}
\chi^A_\bmk(\omega)= [\chi^R_\bmk(\omega)]^*; \qquad 
\chi^K_\bmk(\omega)= 2 \ui \, {\rm Im} \chi^R_\bmk(\omega)
\coth(\tfrac{\hbar \omega}{2 \kB T} ). 
\label{Eq:chi-eq01} 
\tag{S4}
\end{equation}
Here the retarded component of $\check{\chi}_{\bmk}(\omega)$ is given by 
$\chi^R_\bmk(\omega)
= \chi_N/(1+ \lambda_{\rm sf}^2 k^2 - \ui \omega \tau_{\rm sf})$~[S3] 
where $\chi_N$, $\lambda_{\rm sf}$, and $\tau_{\rm sf}$ are 
the paramagnetic susceptibility, 
the spin diffusion length, and spin relaxation time.

We now consider the phonon-dragged SSE [the process $P_2$ shown in FIG.~2 (b)]. 
The renormalized magnon propagator $\delta \check{X}_\bmq(\omega)$ 
in the present case is given by 
\begin{equation}
  \delta \check{X}_\bmq(\omega) = 
  \check{X}_\bmq(\omega)  \check{\Sigma}_\bmq(\omega) \check{X}_\bmq(\omega), 
  \label{Eq:deltaX01} 
  \tag{S5}
\end{equation}
where 
$\check{X}_\bmq(\omega) 
=
\left( { X_\bmq^R (\omega), \atop 0,} 
     { X_\bmq^K (\omega) 
       \atop X_\bmq^A (\omega)} 
     \right) $
is the bare magnon propagator satisfying the equilibrium condition: 
\begin{equation}
X^A_\bmq(\omega) = [X^R_\bmq(\omega)]^*; \qquad 
X^K_\bmq(\omega) = 2 \ui \, {\rm Im} X^R_\bmq(\omega) 
\coth(\tfrac{\hbar \omega}{2 \kB T}). 
\label{Eq:X-eq01} 
\tag{S6}
\end{equation} 
Here, the retarded component is given by 
$X^R_\bmq(\omega)= (\omega-\widetilde{\omega}_\bmq+ \ui \alpha \omega)^{-1}$, 
where $\widetilde{\omega}_\bmq = \gamma H_0 + \omega_\bmq$ 
is the magnon frequency for uniform mode $\gamma H_0$ 
and exchange mode $\omega_\bmq = D_{\rm ex}q^2/\hbar$. 
In Eq.~(\ref{Eq:deltaX01}), the selfenergy $\check{\Sigma}$ due to phonons 
is given by 
\begin{align}
  \check{\Sigma}_\bmq (\omega) &= 
  \frac{\ui}{2N_F} \sum_{\bmK} \left( \frac{\Gamma_{{\bmK},\bmq}}{\hbar} \right)^2 
  \int_\nu \Big\{ 
  \delta D^R(\nu) \check{X}_{\bmq_-}(\omega_-) \check{\tau}_1 \nonumber \\
  & +  \delta D^A(\nu) \check{\tau}_1  \check{X}_{\bmq_-}(\omega_-) 
  + \delta D^K(\nu) \check{X}_{\bmq_-}(\omega_-)   
  \Big\}, 
  \label{Eq:selfenergy01}
  \tag{S7} 
\end{align}
where 
$\Gamma_{{\bmK},\bmq} = 
\widetilde{g} 
\omega_\bmq 
\sqrt{\frac{ \hbar \nu_\bmK}{2M_{\rm ion}v_{\rm ph}^2}} $ 
is the magnon-phonon interaction vertex with $\nu_\bmK$, 
$v_{\rm ph}$ and $M_{\rm ion}$ being the phonon frequency, 
phonon velocity and the ion mass, 
$\check{{\bm \tau}}$ is the Pauli matrix in the Keldysh space, 
and we have introduced the shorthand notations 
$\omega_-= \omega-\nu$, $\bmq_-= \bmq- \bmK$, and 
$\int_\nu = \int_{-\infty}^\infty \frac{d \nu}{2 \pi}$. 
In Eq.~(\ref{Eq:selfenergy01}), the full phonon propagator 
$\delta \widehat{D}_\bmK$ 
[the whole of the phonon lines for $P_2$ in FIG.~2 (b)] 
is written as~[S4] 
\begin{equation}
  \delta \widehat{D}_\bmK (\nu) = 
  \delta \widehat{D}^{l \mathchar`-eq}_\bmK (\nu) 
  + \delta \widehat{D}^{n \mathchar`-eq}_\bmK (\nu). 
  \label{Eq:Dnonloc01}  
  \tag{S8}
\end{equation}
Here, 
$\delta \widehat{D}^{l \mathchar`-eq}(\omega) 
=
\left( { \delta D^{l \mathchar`-eq,R} (\nu), \atop 0,} 
     { \delta D^{l \mathchar`-eq,K} (\nu) 
       \atop \delta D^{l \mathchar`-eq,A} (\nu)} 
     \right) 
$
is the local-equilibrium 
propagator satisfying the local-equilibrium conditions 
$\delta {D}_\bmK^{l \mathchar`-eq,A}(\nu) 
= [\delta {D}_\bmK^{l \mathchar`-eq,R}(\nu)]^*$ and 
$\delta {D}_\bmK^{l \mathchar`-eq,K}(\nu) = 
[\delta {D}_\bmK^{l \mathchar`-eq,R}(\nu)-\delta {D}_\bmK^{l \mathchar`-eq,A}(\nu)] 
\coth(\frac{\hbar \nu}{2 \kB T})$ with its retarded component given by 
\begin{equation}
  \delta {D}_\bmK^{l \mathchar`-eq,R} (\nu) 
  = 
  \frac{\Omega_0^2}{N_F (\Lambda/a_S)} \sum_{\bmK'}  
  \big[{D}^R_\bmK(\nu) \big]^2  {D}^R_{\bmK'}(\nu), 
  \tag{S9}
\end{equation}
while $\delta \widehat{D}^{n \mathchar`-eq} = 
({0, \atop 0,} 
{ {\delta {D}^{n \mathchar`-eq,K} } \atop 0})$ 
is the nonequilibrium propagator with 
its Keldysh component given by 
\begin{align}
  \delta {D}_\bmK^{n \mathchar`-eq,K} (\nu) &= 
  \frac{\Omega_0^2}{N_F (\Lambda/a_S)} 
  \sum_{\bmK'}  
  [{D}^R_{\bmK'}(\nu)-{D}^A_{\bmK'}(\nu)] 
  |{D}^R_\bmK(\nu)|^2  \nonumber \\
  & \times 
  \big[ \coth(\tfrac{\hbar \nu}{2 \kB T_{F_2}}) 
  - \coth(\tfrac{\hbar \nu}{2 \kB T_{{F_1}}}) \big], 
  \label{Eq:dD_noneq01}
  \tag{S10}
\end{align}
In the above equations, $\Omega_0=\sqrt{K_{\rm ph}/M_{\rm ion}}$ 
with the elastic constant $K_{\rm ph}$ in $F$, and 
$\widehat{D}_\bmK(\nu) 
=
\left( { D_\bmK^R (\nu), \atop 0,} 
     { D_\bmK^K (\nu) 
       \atop D_\bmK^A (\nu)} 
     \right) $
is the bare phonon propagator satisfying the equilibrium condition: 
\begin{equation}
D^A_\bmK(\nu) = [D^R_\bmK(\nu)]^*; \qquad 
D^K_\bmK(\nu) = 2 \ui \, {\rm Im} D^R_\bmK(\nu) 
\coth(\tfrac{\hbar \nu}{2 \kB T}), \label{Eq:D-eq01} 
\tag{S11}
\end{equation}
with its retarded component and the phonon lifetime given by 
$D^R_\bmK(\nu)= 
(\nu-\nu_\bmK + \ui /\tau_{\rm ph})^{-1}- 
(\nu+\nu_\bmK+ \ui/\tau_{\rm ph})^{-1}$ 
and $\tau_{\rm ph}$. 

Now substituting these expressions into equation~(\ref{Eq:deltaX01}) 
and use Eq.~(\ref{Eq:I_s01}), the spin current 
injected into $N_1$ by the phonon-drag process is calculated as 
\begin{align}
  I_s^{\rm drag}(N_1) &= 
  \frac{R}{N_N N_F}
  \sum_{\bmk,\bmq,\bmK} 
  \int d \nu_\bmK \nu^4_\bmK \big(\Gamma_{\bmK,\bmq}\big)^2 
  A_{\bmk,\bmq}(\nu) \nonumber \\
  & \times \big[ \coth(\tfrac{\hbar \nu_\bmK}{2 \kB T_{2}}) 
    - \coth(\tfrac{\hbar \nu_\bmK}{2 \kB T_{1}}) \big], 
\label{Eq:ph-drag_main01}
\tag{S12}
\end{align}
where $R= \sqrt{2} (J^2_{sd}S_0) \Omega_0^2 N_{\rm int} (a_S/\Lambda) 
\tau_{\rm ph} /(4 \pi^3 \hbar^4 \nu_D^6 ) $ with $\nu_D= v_{\rm ph}/a_S$, and 
\begin{align}
  A_{\bmk,\bmq}(\nu) &= 
  \int_\omega {\rm Im}\chi^R_\bmk (\omega) {\rm Im}X^R_{\bmq_-}(\omega_-) 
  |X^R_\bmq(\omega)|^2 \nonumber \\
& \times      [\coth(\tfrac{\hbar \omega_-}{2 \kB T_1}) 
        - \coth(\tfrac{\hbar \omega}{2 \kB T_1})]. 
      \tag{S13}
\end{align}
By setting $T=T_2$, $\Delta T= T_1-T_2$, $\Omega_0 = \nu_D$ 
and after some algebra, 
we obtain Eq.~(3) in the main text.

\subsection*{References}

\begin{description}
\item [$S1$] H. Adachi, J. Ohe, S. Takahashi, and S. Maekawa, 
  Phys. Rev. B (in press); arXiv:1010.2325. 

\item [$S2$] A. I. Larkin and Yu. N. Ovchinnikov, 
{Zh. Eksp. Teor. Fiz.} {\bf 68}, 1915 (1975) 
[Sov. Phys. JETP {\bf 41}, 960 (1975)]. 

\item [$S3$] P. Fulde and A. Luther, 
Phys. Rev. {\bf 175}, 337 (1968). 

\item [$S4$] K. Michaeli and A. M. Finkel'stein, 
Phys. Rev. B {\bf 80}, 115111 (2009). 

\end{description}

\end{document}